# RAMAN PHONONS AND AGEING-RELATED SURFACE DISORDER IN $Na_xCoO_2$

M. N. Iliev, A. P. Litvinchuk, R. L. Meng, Y. Y. Sun, J. Cmaidalka, C. W. Chu

*Texas Center for Superconductivity and Advanced Materials*

*and Physics Department, University of Houston, Houston TX 77204-5002, USA*

**Abstract**

The polarized Raman spectra from *ab* and *ac* surfaces of single crystal $Na_xCoO_2$ (x≈0.7), parent compound of recently discovered superconductor $Na_xCoO_2 \cdot yH_2O$, are reported and discussed. The crystals were hexagon platelets of typical size 3×3×0.1 mm. Three of the five ($A_{1g}+E_{1g}+3E_{2g}$) Raman active phonons were unambiguously identified at 458 ($E_{1g}$), 494 ($E_{2g}$) and 574 ($A_{1g}$) cm$^{-1}$. The spectra from *ab* and *ac* surfaces differ significantly and provide evidence that within hours after preparation the *ac* surface, unlike the *ab* one, is strongly disordered. Within several days the disorder extends over the *ab* surface too.

$Na_xCoO_2$ has a hexagonal structure (space group #194, *P6₃/mmc*, Z=2) consisting of $CoO_2$ and Na layers parallel to the *ab* planes (Fig.1). It attracted significant interest after superconductivity at 5K was recently reported for the closely related $Na_xCoO_2 \cdot yH_2O$ (x≈0.35, y≈1.3), thus providing evidence that superconductivity may occur in two-dimensional $CoO_2$ layers [1-3]. As the pairing through phonons is typically the prime candidate for explaining a superconductivity mechanism, the detail knowledge of phonon structure of superconducting materials and their parent compounds is of definite importance. Here we present the polarized Raman spectra of $Na_xCoO_2$ (x≈0.7) and discuss the assignment of the observed Raman phonon lines, as well as the structural disorder developing in hours and days time scale, starting from the *ac* surface.

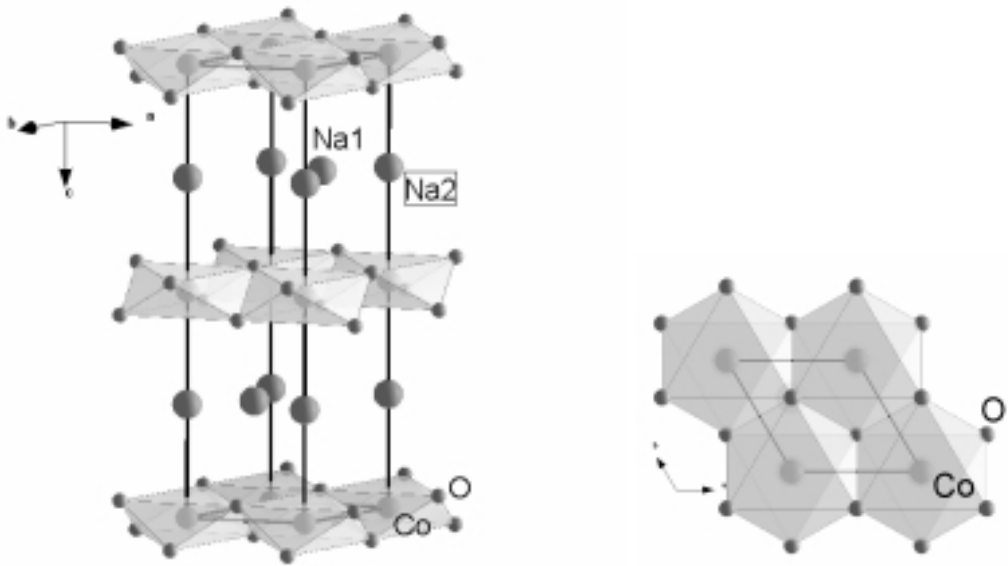

Fig.1 Elementary cell of $Na_xCoO_2$ (x =1)

Single crystals of $Na_xCoO_2$ (x≈0.7) were grown by the flux method. High purity $Na_2CO_3$ (>99.9%) and $Co_3O_4$ ( >99.99%) powders were used as starting material and NaCl as a flux. They were mixed throughout in a molar ratio of Na:Co:NaCl = 1:1:10 and placed in an $Al_2O_3$ crucible. The crucible was heated to 950°C for 12 hours in air, and then cooled down to 800°C at a rate of 0.5°C/h. The molten material was than washed by distill water to remove the NaCl flux. The single crystals of $Na_xCoO_2$ have the form of hexagon-shape thin platelets of typical size 3×3×0.1 mm (Fig2). A visual observation under microscope reveals that each crystal platelet is in fact a pack of thinner layers (d < 5 μm) parallel to the hexagon surfaces. The lattice parameters $a$ = 2.82 Å, $c$ = 10.92 Å, as determined from X-ray powder diffraction, are close to those reported in Ref.[1].

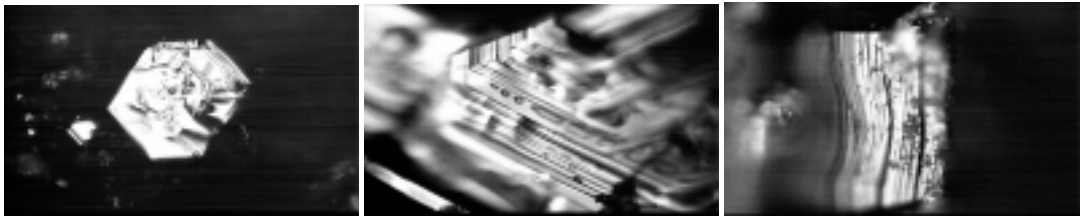

Fig.2.

Typical $Na_{0.7}CoO_2$ single crystals consisting of thin layers packeted parallel to the *ab*-plane.

Polarized Raman spectra were measured at room temperature from *ab* and *ac* crystal surfaces, using a microscope with 50× or 100× objective to focus the incident laser beam (514.5 nm or 632.8 nm) at a spot of 2-3 μm diameter and collect the scattered light. The following exact scattering configurations were available: *xx*, *yy*, *zz*, *xy*, *zz*, and *zx*. The first and second letters in these notations stay for the polarization directions of the inci-dent and scattered light, respectively, *x* and *y* are two orthogonal directions in the *ab* plane, *z* is parallel to the *c*-axis.

In Table I are summarized the atomic site positions, site symmetries and the Γ-point (zone center) phonon modes expected for NaCoO$_2$. Five modes ($A_{1g}+E_{1g}+3E_{2g}$) are Raman active. The $A_{1g}$ and $E_{1g}$ modes involve motions of the only oxygen atoms. Both Na and O motions may participate in the $E_{2g}$ modes. Co motions are not Raman active.

Fig. 3a shows the polarized Raman spectra of Na$_{0.7}$CoO$_2$ as obtained on the next day after the samples were prepared. The lower two spectra (*xx* and *xy*) measured from the *ab* surface allow to unambiguously identify the $A_{1g}$ (574 cm$^{-1}$) and $E_{1g}$ (458 cm$^{-1}$) lines. These frequencies are in excellent agreement with the ones predicted by calculations of lattice dynamics [4]. The line at 494 cm$^{-1}$ in the *zx* spectrum obviously corresponds to one of the three $E_{2g}$ modes. Surprisingly, in the upper *xx* and *zz* spectra, measured from the *ac* surface, the Raman bands corresponding to the $A_{1g}$ and $E_{1g}$ modes are much broader and shifted to ≈ 600 cm$^{-1}$ and 480 cm$^{-1}$, respectively. An unusually broad band can also be seen in the *zx* spectrum, which was taken from the *ac* plane too.

Table I. Atomic positions, site symmetries and Raman modes in $Na_xCoO_2$ (x =1)

| Atomic position | Wickof index | Site symmetry | Irreducible representations | Raman modes |
|---|---|---|---|---|
| Co (0,0,0) | 2a | $D_{3d}$ | $A_{2u}+B_{2u}+E_{1u}+E_{2u}$ | |
| O (⅓,⅔,z) | 4f | $C_{3v}^d$ | $A_{1g}+A_{2u}+B_{1g}+B_{2u}+E_{1g}+E_{1u}+E_{2g}+E_{2u}$ | $A_{1g}+E_{1g}+E_{2g}$ |
| Na1 (⅔,⅓,¼) | 2d | $D_{3h}^1$ | $A_{2u}+B_{1g}+E_{1u}+E_{2g}$ | $E_{2g}$ |
| Na2 (0,0,¼) | 2b | $D_{3h}^1$ | $A_{2u}+B_{1g}+E_{1u}+E_{2g}$ | $E_{2g}$ |
| **Non-zero elements of the Raman tensors** ||||| 
| $A_{1g} \rightarrow \alpha_{xx}, \alpha_{yy} = \alpha_{xx}, \alpha_{zz}$ ||||| 
| $E_{1g} \rightarrow \alpha_{xx}, \alpha_{yy} = -\alpha_{xx}; \alpha_{xy}$ ||||| 
| $E_{2g} \rightarrow \alpha_{zx} = \alpha_{xz}; \alpha_{yz} = \alpha_{zy}$ ||||| 

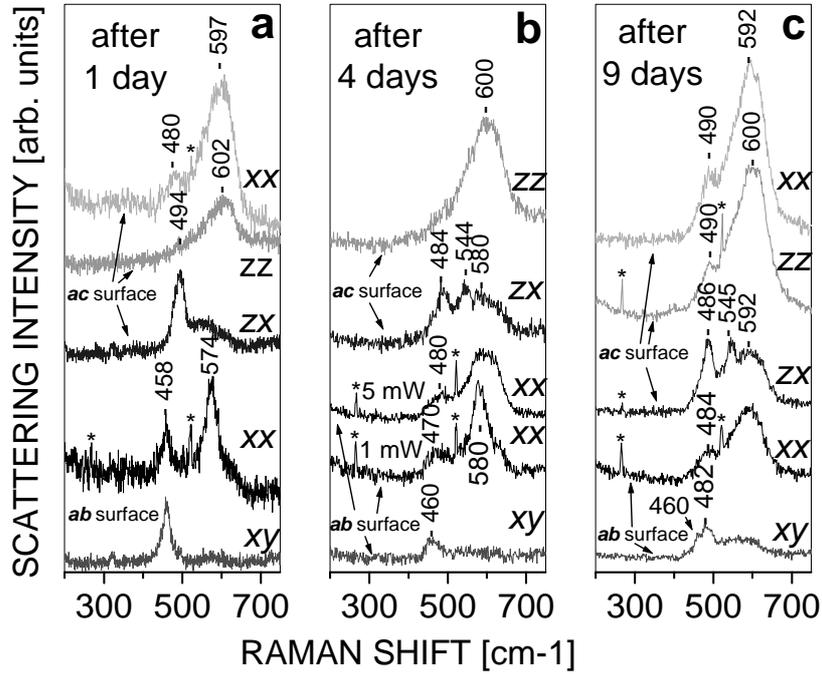

Fig.3. Polarized Raman spectra of $Na_xCoO_2$ (x≈0.7) measured (a) one day, (b) four days, and (c) nine day after sample preparation with 514.5 nm excitation. Asterisks mark the laser plasma lines.

The drastic difference between the spectra from *ab* and *ac* surfaces is a strong indication that the crystal structure within the depth of the scattering volume (d < 1 μm) below the two surface types is also different. Broad spectra like those from the *ac* surface may be expected for a strongly disordered structure. Indeed, the loss of translational symmetry results in activation of all phonon modes (or all modes related to disordered sublattice). The spectral profile of such disorder-induced Raman scattering qualitatively reflects the smeared phonon density-of-states versus phonon energy distribution for the perfect crystal [5,6]. In the intermediate case both broadened Γ-point phonon lines and bands of density-of-states origin may be expected.

Within plausible assumption that the band broadening and spectral weight redistribution are due to lattice disorder, Figs. 3b and 3c provide evidence that with ageing the disorder increases also near the *ab* surface. On the ninth day after crystal preparation the *xx* spectra from the *ab* and *ac* surfaces (see Fig.3c) become practically identical. It is worth mentioning that order-disorder transformation at the *ab* surface can also be induced by sample heating. The latter is illustrated in Fig.3b where are compared the profiles of two *xx* spectra obtained from the same *ab* surface with different laser power (1 and 5 mW). In some spectra a relatively narrow line is also seen at 545 cm$^{-1}$. As its appearance is spot-dependent, we tentatively assign to a minority phase.

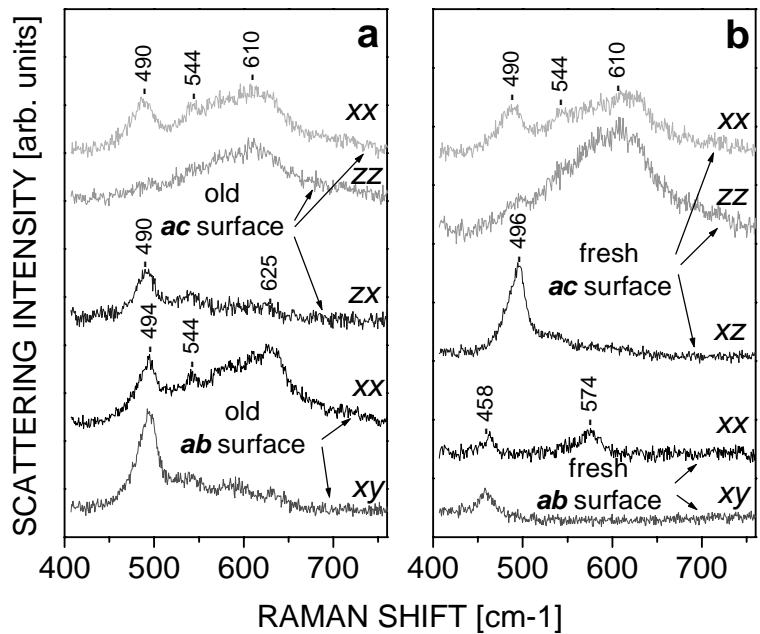

Fig.4. Polarized Raman spectra of $Na_xCoO_2$ (x≈0.7) measured 90 days after sample preparation from: (a) surfaces aged in air; (b) freshly cleaved surfaces. The 632.8 nm laser line was used for excitation

Fig.4a shows the spectra of $Na_{0.7}CoO_2$ crystals obtained from *ab* and *ac* surfaces aged for 90 days in air. The $E_{1g}$ and $E_{2g}$ lines are now shifted to 494 cm$^{-1}$ and 490 cm$^{-1}$, respectively, where the $A_{1g}$ line cannot be distinguished. These three lines, however, are observed at their original positions in spectra are taken from freshly cleaved surfaces (Fig.4b). This clearly indicates that ageing-related structural rearrangement is restricted to the surface layer, not extending in the crystal volume. This is also consistent with the results of x-ray characterization, which provides no evidence for change of the lattice parameters with ageing. Interestingly, although number of freshly cleaved *ac* surfaces were probed, the *xx* and *zz* spectra were always dominated by the broad structure between 500 and 650 cm$^{-1}$ and the $E_{1g}$ mode was observed at ~490 cm$^{-1}$. One can therefore conclude that, unlike *ab* surface, structural rearrangement at the ac surface accomplishes within a short time interval.

The nature of above described time- and temperature-dependent order-disorder transformation at the surface of nominally water-free $Na_xCoO_2$ is not clear at present. Among the factors that may contribute to surface degradation are the deintercalation of sodium, hydration, change of oxygen content and formation of minority phases. The elucidation of their role may be of key importance for understanding the superconductivity in closely related $Na_xCoO_2 \cdot yH_2O$.

**Acknowledgements**

This work was supported in part by the State of Texas through the Texas Center for Superconductivity and Advanced materials, NSF Grant No. DMR-9804325, and the Division of Materials Sciences of the U.S. Department of Energy under Contract No. DE-AC03-76SF00098.